\documentclass[a4paper,10pt]{article}
\usepackage{
times,
amsmath,amssymb,epsfig,graphics,graphicx}

\newcommand{\e}{\textrm { e}}

\newcommand{\be}{\begin{eqnarray}}
\newcommand{\ee}{\end{eqnarray}}

\begin{document}

\begin{center}
{\Large\bf The Quantum Mechanics Problem of
the  Schr\"odinger Equation with the 
Trigonometric Rosen-Morse Potential}
\end{center}
\vspace{0.5cm}

\begin{center}
{\Large C.\ B.\ Compean, M.\ Kirchbach}
\end{center}

\vspace{0.5cm}
\begin{center}
{\it Instituto de F\'{\i}sica}, \\
         {\it Universidad Aut\'onoma de San Luis Potos\'{\i},}\\
         {\it Av. Manuel Nava 6, San Luis Potos\'{\i}, S.L.P. 78290, M\'exico}
\end{center}

\vspace{0.5cm}

\begin{flushleft}
{\bf Abstract:}
We present the  quantum mechanics problem
of the one-dimensional Schr\"odinger equation with
the trigonometric Rosen-Morse potential. This potential is of possible
interest to quark physics  in so far  as it captures the essentials of the
QCD quark-gluon dynamics and
(i) interpolates between a Coulomb-like
potential (associated with one-gluon exchange) and the infinite wall
potential (associated with asymptotic freedom), 
(ii) reproduces in the intermediary region  the linear
confinement  potential (associated with multi-gluon self-interactions)
as established by lattice QCD calculations of hadron properties.
Moreover, its exact real solutions given here display  a new class of
real orthogonal polynomials and thereby interesting mathematical
entities in their own. 

\end{flushleft}

\vspace{0.3cm}
\begin{flushleft}
\end{flushleft}

\section{Introduction}
There are few  quantum mechanic problems on bound states wave functions
that allow for exact solutions. The examples 
frequented in the standard textbooks on quantum
mechanics,  mathematical methods,  and problem samplers
\cite{textbook}
range from the simplest case of the infinite square well potential
over the more advanced Harmonic-Oscillator-- and Coulomb potentials
and culminate with the more sophisticated Morse--, Eckart--, 
the trigonometric Scarf and the  hyperbolic Rosen-Morse potentials, 
which have importance in atomic and molecular spectroscopy.
All the known exactly soluble potentials are attached to observed physical
spectra and the corresponding wave functions are expressed in terms
of the well known classical orthogonal polynomials \cite{handbook}.

Within the context of supersymmetric quantum mechanics, one encounters
two more potentials, the hyperbolic Scarf and the trigonometric
Rosen-Morse potentials, that are claimed  to allow for exact solutions
however in terms of Jacobi polynomials with complex arguments and 
complex indices \cite{Sukumar}, \cite{Khare} 
a less appealing feature indeed given the observation on the 
non-trivial orthogonality
properties of the complex Jacobi polynomials as reported in the current
 mathematical literature \cite{Jacobi-c}, \cite{Jacobi-otros}.

In our previous work \cite{CK} we demonstrated within the context
of supersymmetric quantum mechanics that the one-dimensional
Schr\"odinger equation with the trigonometric Rosen-Morse
potential is exactly soluble in terms of real orthogonal polynomials 
of a new type. Here we work out  these solutions anew within the different 
context of interpolation between the Coulomb and the infinite square well
potentials and hint on  the possible importance of the trigonometric
Rosen-Morse potential for QCD.
  
The paper is organized as follows. In the next section we analyze the 
shape of the trigonometric Rosen-Morse potential.
In section III we present the exact real orthogonal polynomial 
solutions of the corresponding Schr\"odinger equation. The paper
ends with a brief concluding section.

\section{The shape of the trigonometric Rosen-Morse potential }

We adopt the following form
of the trigonometric Rosen-Morse potential \cite{Sukumar},\cite{Khare} 
\begin{equation}
 v(z)=-2 b \cot z +a(a+1)\csc^2 z\, , 
\label{v-RMt}
\end{equation}
with $a>-1/2$ and displayed in  Fig.~1.
Here,  
\begin{eqnarray}
z=\frac{y}{d},&\quad& v(z)=V(zd)/(\hbar^{2}/2md^{2}),
\nonumber\\
\epsilon_n &=&E_n/ (\hbar^{2}/2md^{2})\,,
\end{eqnarray}
with $y$ being the one-dimensional variable, $ d$ 
a properly chosen length scale, $V(y)$ the potential in 
ordinary coordinate space, and $E_n$ the energy level.

\noindent
Our point here is that $v(z)$ interpolates between the Coulomb and
the infinite wall potential going through an intermediary region of
linear $z$ and harmonic-oscillator $z^2$ 
dependences. To see this (besides inspection of
Fig.~1) it is quite instructive to expand the potential
in a Taylor series which for appropriately small $z$,
takes the form of a Coulomb-like 
potential with a centrifugal-barrier like term
provided by the $\csc^2 z$ part,
\begin{eqnarray}
v(z)\approx -\frac{2b}{z} +\frac{a(a+1)}{z^2}\, .
\label{Taylor_Coul}
\end{eqnarray}
In an intermediary range where inverse powers of $z$ may be neglected,
one finds the linear plus a harmonic-oscillator potentials
\begin{equation}
v(z)\approx
\frac{2b}{3}\, z +\frac{a(a+1)}{36}z^2\, .
\label{Taylor_lin_HO}
\end{equation}
Finally, as long as  $\cot\, z \stackrel{z\to \pi}{\longrightarrow}
\infty$ and $\csc^2\, z \stackrel{z\to \pi}{\longrightarrow}\infty$,
the potential obviously evolves to an infinite wall.
Below we shall show that  in the parameter limit
$a\to 0$ and $b\to 0$, the wave functions recover those of
the infinite square wall.

\begin{figure}
\begin{center}
\includegraphics[width=70mm,height=70mm]{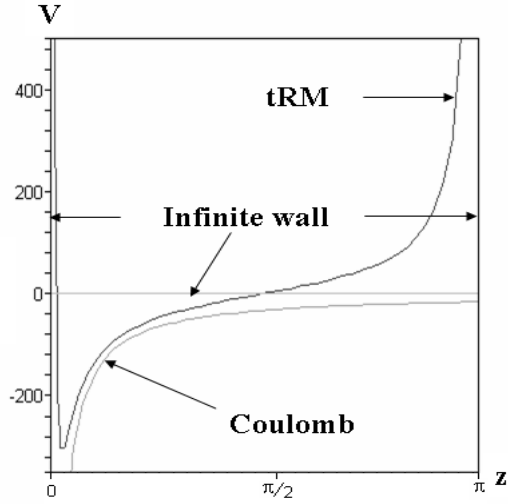}
\caption{ The trigonometric Rosen-Morse potential and its
proximity to the Coulomb-like and the infinite square wall potentials for
$a=0.25$ and $b=1$.}
\end{center}
\end{figure}

Above shape captures surprisingly well the essentials of the QCD 
quark-gluon dynamics where the one gluon exchange gives rise to 
an effective Coulomb-like potential, while the self-gluon interactions
produce a linear potential as established  by lattice calculations
of hadron properties.
Finally, the infinite wall piece of the trigonometric
Rosen-Morse potential provides the regime suited for  
asymptotically free quarks.

\noindent
By above considerations one is tempted to conclude that 
the potential under consideration may be a good candidate for an effective
QCD potential.

\section{Exact spectrum and wave functions of the trigonometric Rosen-Morse 
potential}

The one-dimensional  Schr\"odinger equation with the trigonometric
Rosen-Morse potential (tRM) reads:
\be {d^{\, 2}\ R_m(z) \over d\ z^2}+\left(
2 b\cot z-a(a+1)\csc^2 z+\epsilon\right) R_m(z)=0\, ,
\label{Sch-RMt}\ee

Our pursued strategy in solving it 
will be to first reshape it to
the particular case of a  self-adjoined Sturm-Liouville 
equation of the form
\be s (x){{d^{\, 2}F_m(x)} \over {d\ x^2}}+
{1\over {w(x)}}\left({{d\ s(x)w(x)}\over {d\ x}}\right){d\ F_m(x)
\over d\ x}+\lambda_m \ F_m(x)=0\ ,\label{d2-R2}\ee
and then try to solve it by means of  the so called
Rodrigues representation 
\be F_m(x)=\frac{1}{K_mw(x)}{d^m\over d\ x^m}(w(x)\ s(x)^m) \ ,
\label{Rodrigues-0}\ee
where $K_m$ is the  normalization constant of the $F_m(x)$ polynomials.
The constant $\lambda_m$ in Eq.~(\ref{d2-R2}) is
supposed to satisfy the following condition \cite{Dennery}
\begin{eqnarray}
\lambda_m&=&-m\left(K_1{{d\ F_1(x)} \over {d\ x}}+
{1\over 2}(m-1) {{d^{\, 2} s(x)}\over 
{d\ x^2}}\right)\, ,\quad \mbox{with}\nonumber\\ 
 F_1 (x) &=& {1\over {K_1w(x)}}
\left({{d\ s(x)w(x)}\over {d\ x}}\right)\, .
\label{lamb}
\end{eqnarray} 
Sturm-Liouville equations of the type given in Eq.~(\ref{d2-R2})
are called {\it hypergeometric\/} equations.

The chosen strategy is inspired by the observation that
all the classical polynomials have been  obtained precisely from 
those very Eqs.~(\ref{d2-R2})--(\ref{lamb}), appear
orthogonal with respect to the weight function $w(x)$
and obey the following restrictions 
(see Chpt.~10 in Ref.~\cite{Dennery} for more details):
(i) $F_1(x)$ is a polynomial of first order,
(ii) $s(x)$ is a polynomial of at most second order and  real
roots,
(iii) $w(x)$ is real, positive and  integrable within a given interval
 $[a,b]$, and satisfies the boundary conditions
\be w(a)s(a)= w(b)s(b)=0 \ .
\label{rule_tobreak}\ee

\begin{quote}
We here draw attention to the fact that the exact solutions of 
Eq.~(\ref{Sch-RMt}) can be expressed in terms of real orthogonal 
polynomials that solve a hypergeometric differential
equation of a new class.
\end{quote}

Back to Eq.~(\ref{Sch-RMt})  we factorize the wave function as
\be R_m(z) =\e^{-\alpha z/2} (1+\cot ^2 z)^{\frac{-(1-\beta )}{2}}
C^{(\alpha , \beta )}_m(\cot z)\, , 
\label{Schroed}
\ee
with $\alpha$ and $\beta$ being constants.
Upon introducing the new variable $x=\cot z$ and
substituting the above factorization ansatz
into Eq.~(\ref{Schroed}) and a subsequent division by $(1+x^2)^{(1-
\beta)/2}$ one finds the equation
\begin{eqnarray}
&&(1+x^2)
\frac{d^{\, 2}\ C^{(\beta, \alpha) }_m(x)}{d\ x^2}+
2\left({\alpha\over 2}+\beta x\right) 
{d\ C^{(\beta ,\alpha ) }_m (x) \over d\ x }\nonumber\\
&+&\left((-\beta(1-\beta)-a(a+1)) +{(-\alpha(1-\beta)+2 b)x + \left(
\left({\alpha\over 2}\right)^2-(1-\beta )^2+\epsilon_n\right)\over 1+x^2}\right)
C^{(\beta ,\alpha )}_m(x)  =  0\ , \label{Sch-RMt4}
\end{eqnarray}
which is suited for comparison with the hypergeometric equation (\ref{d2-R2}).

The derivative terms in Eq.~(\ref{Sch-RMt4}) have already the desired form of 
Eq.~(\ref{d2-R2}).
As a first observation one encounters $s(x)$ as
\be
s(x)=1+x^2\, ,
\label{r_im_r}
\ee
and of purely imaginary roots. Nonetheless, as we shall see immediately,
this is not to turn out to be a great obstacle on the way of finding
the exact real solutions of Eq.~(\ref{Schroed}).
Next, the function that plays the role of the weight-function is
\begin{equation}
w^{(\beta,\alpha)}(x)=
(1+x^2)^{\beta -1}e^{-\alpha\cot ^{-1}x}\, .
\label{weight_funct}
\end{equation}
These $s(x)$ and $w^{(\beta ,\alpha )}(x)$ functions allow to write the 
following {\it new \/} hypergeometric differential equation for 
the  polynomials $C_m^{(\beta ,\alpha )}(x)$:
\begin{equation}
(1+x^2)
\frac{d^{\, 2}\ C^{(\beta, \alpha) }_m(x)}{d\ x^2}+
2\left({\alpha\over 2}+\beta x\right) 
{d\ C^{(\beta ,\alpha ) }_m (x) \over d\ x }\
-m(2\beta +m-1)C_m^{(\beta ,\alpha )}(x)=0\, .
\label{new_pol}
\end{equation}
\begin{quote}
Equation (\ref{new_pol}) generalizes the hypergeometric
equation and represents  a new equation in mathematical physics
that is important on its own.
\end{quote}

As in the case of the Jacobi polynomials, 
also the weight function giving rise to  Eq.~(\ref{new_pol})
happens to be parameter dependent.
If  the potential equation~(\ref{Sch-RMt4}) is to coincide with the
polynomial equation~(\ref{new_pol}) then  the coefficient in front of
the $1/(x^2+1)$ term in (\ref{Sch-RMt4}) has to nullify. This imposes 
the following conditions on the indices of the polynomials 
which are to enter the Schr\"odinger wave function:   
\be
-\alpha(1-\beta)+2 b=0\ ,\label{ab-b-1}\\
\left({\alpha\over 2}\right)^2-(1-\beta)^2+\epsilon=0\label{ep-b2_a2-1}\, .
\ee
With that the Eq.~(\ref{Sch-RMt4}) to which one 
has reduced the original  Schr\"odinger equation  amounts to
\be (1+x^2){d^{\, 2}\ C_m^{(\beta ,\alpha )}(x) \over d\ x^2}+
2\left({\alpha\over 2}+\beta x\right) 
{d\ C_m^{(\beta ,\alpha )}(x)
\over d\ x} + (-\beta(1-\beta)-a(a+1))C_m^{(\beta ,\alpha )}(x) = 0\ . 
\label{Sch-RMt5}\ee

The final step is to identify the constant term in the latter equation with
the one in Eq.~(\ref{new_pol}) which leads to a third  condition
\be -\beta(1-\beta)-a(a+1) = -m(2\beta+m-1)\ .\label{b-1}\ee
Remarkably, Eqs.~(\ref{ab-b-1}), (\ref{ep-b2_a2-1}) and (\ref{b-1})
indeed allow for consistent  solutions for $\alpha$, $\beta $, and $\epsilon$
and given by (upon renaming $m$ by $(n-1))$:

\be
\beta_n=-(n+a)+1\ ,&\quad& \alpha_n={2 b\over n+a}\ ,\nonumber\\
\epsilon_n &=& (n+a)^2-{b^2\over (n+a)^2}\ ,
\label{RMt_spectrum}
\ee
with  $n\ge 1$.
In this way Eq.~(\ref{RMt_spectrum}) provides 
the exact tRM spectrum displayed in Fig.~2.
\begin{figure}
\begin{center}
\includegraphics[width=70mm,height=70mm]{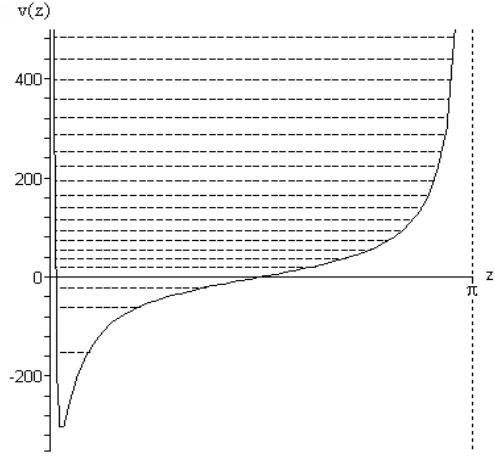}
\caption{ Energy levels within the trigonometric Rosen-Morse potential
for $a=0.25$ and $b=1$.}
\end{center}
\end{figure}
With that all the necessary ingredients for the solution of 
Eq.~(\ref{Sch-RMt5}) have been prepared. In now exploiting the
Rodrigues representation (when making the $n$ dependence explicit)
\be C^{(\beta_n,\alpha_n)}_{n}(x)=
{1\over K_n\ w^{(\beta_n, \alpha_n)}(x)}
{d^{n-1}\over d\ x^{n-1}}\left(w^{(\beta_n,\alpha_n)}(x)\ 
s(x)^{n-1}\right) \, , \label{pol-nvo}\ee
allows for the systematic construction of 
the solutions of Eq.~(\ref{Sch-RMt5}).
To be specific, 
the lowest five $C_n^{(\beta_n,\alpha_n)}(x)$ polynomials that enter
the exact wave function of the Schr\"odinger equation with the
trigonometric Rosen-Morse potential are now obtained as:
\be
C^{(\beta_1,\alpha_1)}_{1}(x)&=&{1\over K_1}\ , \\
C^{(\beta_2,\alpha_2)}_{2}(x)&=&
{2\over K_2}\left(-(1+a)x + {b\over 2+a}\right)\ , \\
C^{(\beta_3,\alpha_3)}_{3}(x)&=&
{2\over K_3}\left((1+a)(2a+3)x^2-2(2a+3){b\over 3+a}x+ 
\left({2b^2\over (3+a)^2}-(1+a)\right)\right)\ , \\
C^{(\beta_4,\alpha_4)}_{4}(x)&=&{4\over K_4}
\left(-(1+a)(2a+3)(2+a)x^3 + 3(a+2)(2a+3){b\over (4+a)} x^2\right. \nonumber 
\\
& &\left.  -3(2+a)\left(2 {b^2\over (4+a)^2}-(1+a) \right)x + \left({2b^3\over 
(4+a)^3}-(3a+4){b \over 4+a}\right) \right) \ ,\\
C^{(\beta_5,\alpha_5)}_{5}(x)&=&{4\over K_5}\left((1+a)(2a+3)(2+a)
(2a+5)x^4 - 4(2a+3) (2+a) 
(2a+5) {b\over (5+a)} x^3 \right.  \nonumber \\
& &\left. +6(2+a) (2a+5) \left({2 b^2\over (5+a)^2}-(1+a) \right)x^2 \right.  
\nonumber \\
& & - 4 (2a+5) \left({2b^3\over (5+a)^3}-(3a+4){b \over 5+a}\right)x \nonumber 
\\
& &\left. +\left({4b^4\over (5+a)^4}-{4b^2\over (5+a)^2}(3a+5)+3(2+a)(1+a)
\right) \right) \ ,
\label{C-pol}
\ee
where $x=\cot z$.\\
Above polynomials solve exactly Eq.~(\ref{Sch-RMt5}) which can be immediately
cross-checked  by back-substituting Eqs.~(\ref{C-pol}) 
into Eq.~(\ref{Sch-RMt5}). Employing symbolic mathematical programs is 
quite useful in that regard.
\footnote{We emphasize that the general weight function 
in Eq.~(\ref{weight_funct})
is only parameter dependent and that it was the Schr\"odinger equation
that gave these parameters the  particular $n$ dependent values in 
Eq.~(\ref{pol-nvo}). 
The claim in Ref.~\cite{CK} that the
$C_n^{(\beta ,\alpha)}(x)$ polynomials require  
an $n$ dependent weight-function restricts to the polynomials
that enter the wave functions to the Schr\"odinger equation 
alone and does not extend  to the general solutions of Eq.~(\ref{new_pol}).
Notice change of the notations from $C_n^{(a,b)}(x)$ in Ref.~\cite{CK}
to $C_n^{(\beta ,\alpha )}(x)$ in the present work. The change has been 
dictated by the necessity to distinguish between the general solutions of
the hypergeometric equation~(\ref{new_pol}) of a new class
and the particular polynomials that define the
solutions of the Schr\"odinger wave equation with the
trigonometric Rosen-Morse potential in which case
the polynomial indices acquire $n$ dependence.}

Notice that in terms of $w^{(\beta_n ,\alpha_n)}(x)$ the wave function 
is expressed as
\be
R_n(\cot^{-1} x)=\sqrt{w^{(\beta_n , \alpha_n)} (x)}
C_n^{(\beta_n ,\alpha_n )}(x)\, .
\label{R_z}
\ee
In Figs.~3 we display as an illustrative example the 
wave functions for the first four (unnormalized) levels.
A version interesting for physical application (see concluding Section) 
is the one with $a=0$ when the potential becomes 
\be
v(z)=-2b\cot \, z\, ,
\label{RVM_pot}
\ee
and in which case the normalization constant can be calculated in the
following closed form: 
\be K_n=\left({(n!)^2 n^3(1-\e^{-2\pi b/n})\over 4 b (b^2+n^4)}\right)^{1/2}\ .
\label{RVM_norm_const}
\ee
The associated energy spectrum is given  by
\begin{equation}
\epsilon_n=n^{2} -\frac{b^{2}}{n^{2}}\, .
\label{RVM_spctr}
\end{equation}
Correspondingly, the  wave functions simplify to
\be
R_n(z)=e^{-\frac{bz}{n}}\sin ^n z C_n^{(-n+1,\alpha_n)}(\cot \, z)\,.
\label{psi6}
\ee

\begin{figure}
\begin{center}
{(a)}\includegraphics[width=70mm,height=70mm]{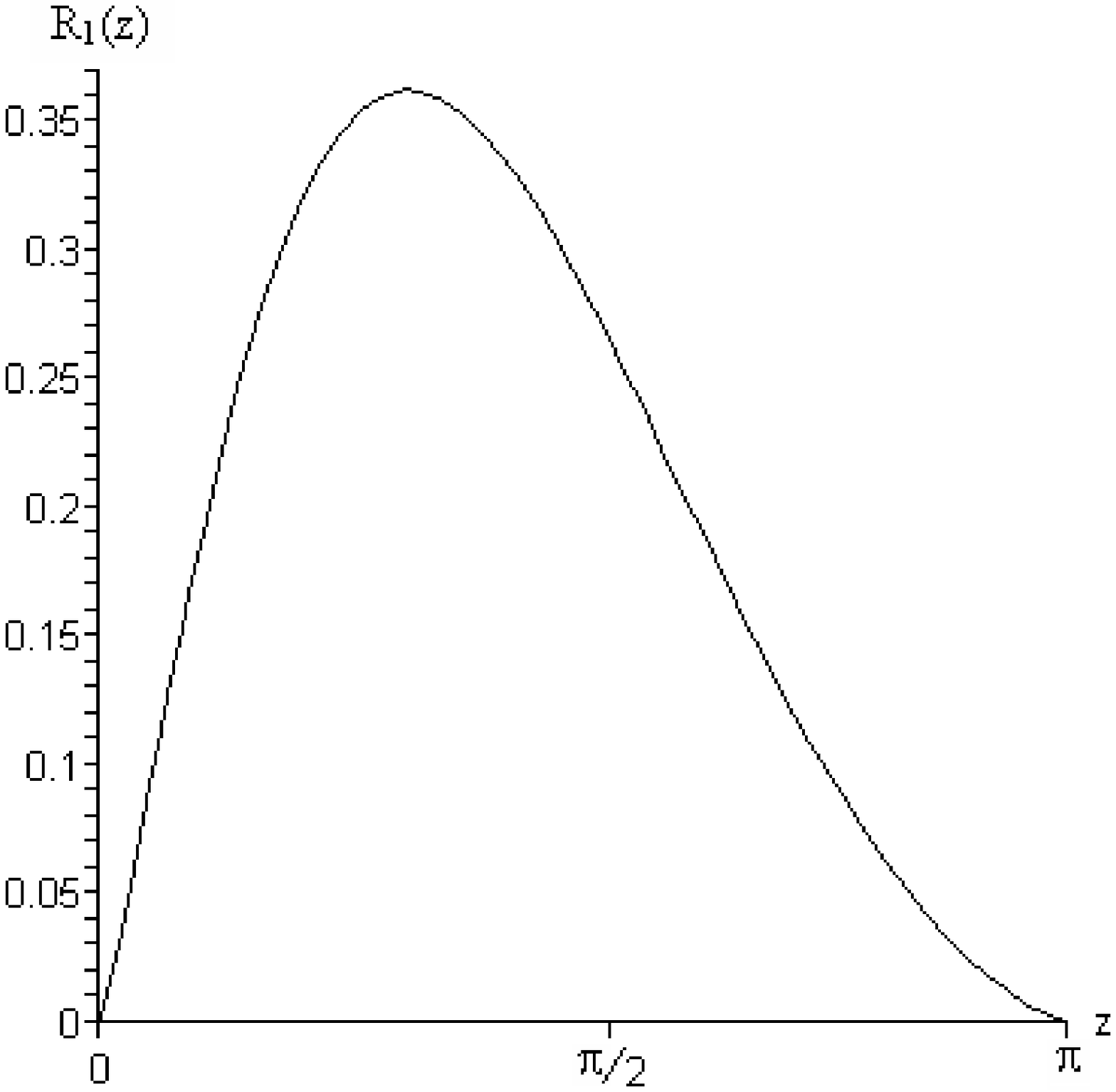}
{(b)}\includegraphics[width=70mm,height=70mm]{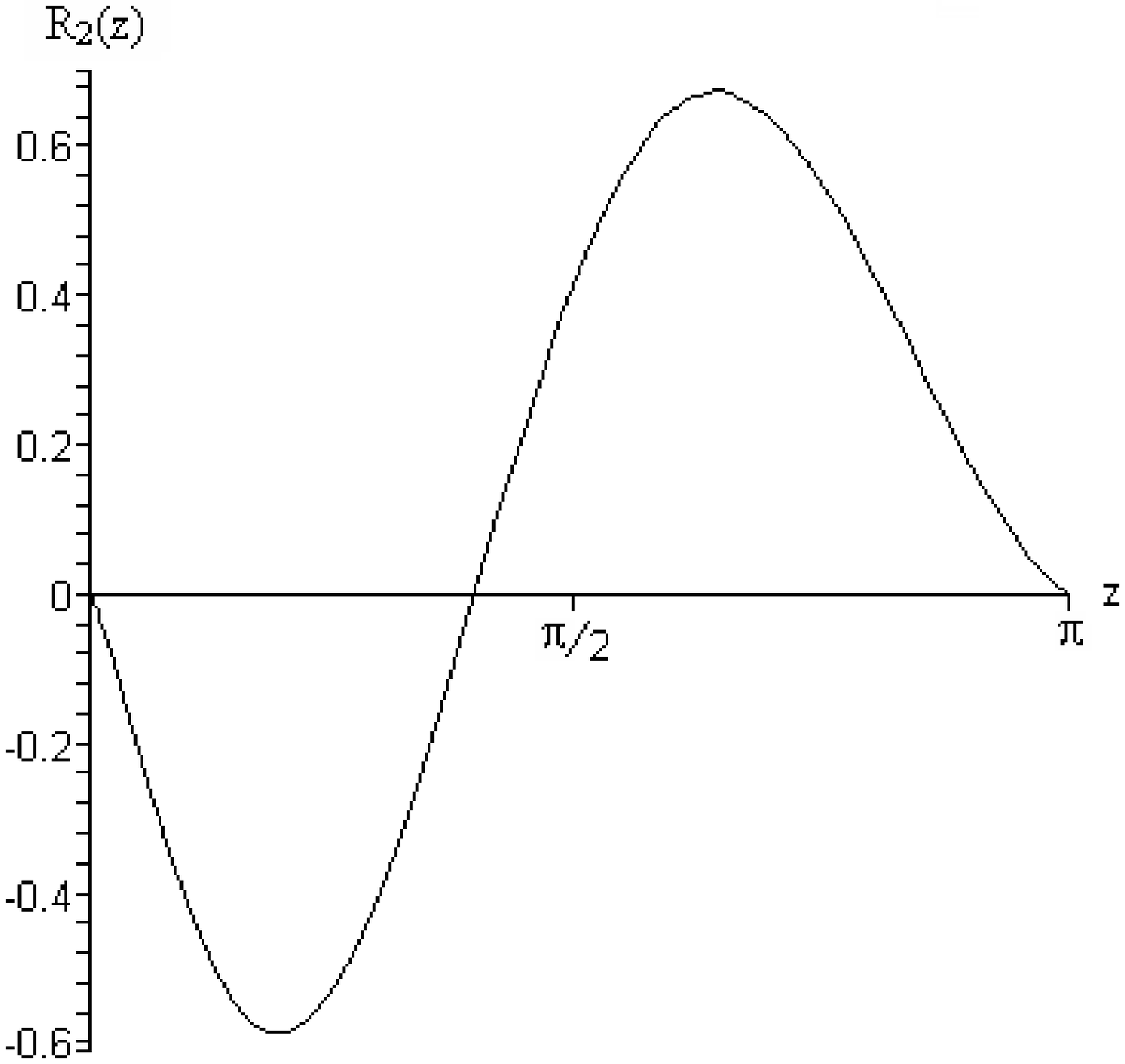}
{(c)}\includegraphics[width=70mm,height=70mm]{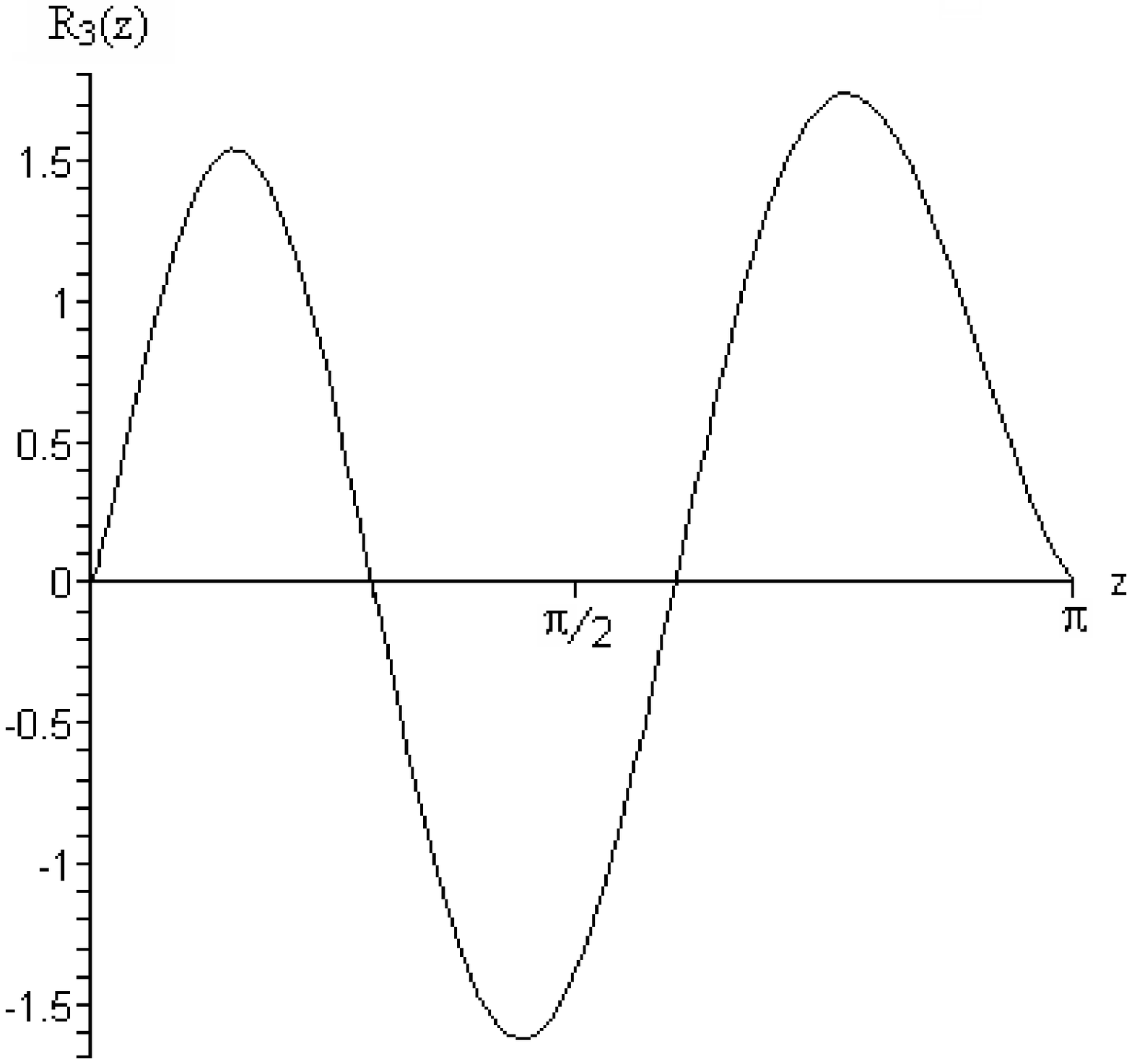}
{(d)}\includegraphics[width=70mm,height=70mm]{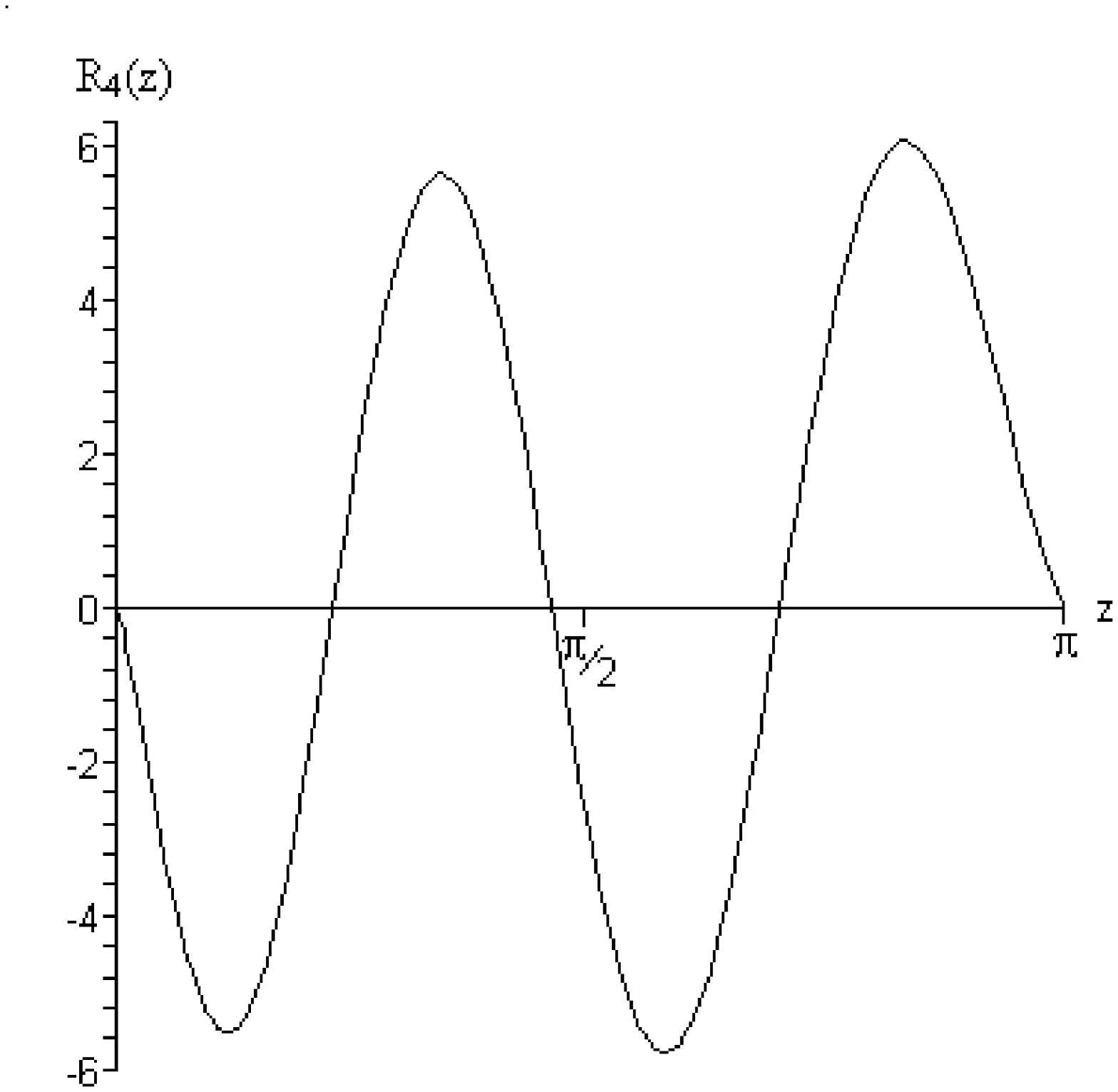}
\end{center}
\caption{ Wave functions for the first four levels in the trigonometric
Rosen-Morse potential  for $a=$0.25 and $b=$1.} 
\end{figure}

\noindent
Next one can check orthogonality of the physical solutions
and obtain it as it should be as
\be \int_0^\pi dz\ R_{n}(z) R_{n'}(z) =\delta_{n\ n'}\, .
\label{orth_R}\ee

The orthogonality of the wave functions $R_n(z)$  implies  in $x$ space 
orthogonality of the $C_n^{(\beta_n,\alpha_n)}(x)$ polynomials with respect to
$w^{(\beta_n, \alpha_n)}(x)\frac{dz}{dx}$ due to the variable change.
As long as $\frac{d \cot^{-1} z}{dx}=-1/(1+x^2)\equiv -1/s(x)$
then the orthogonality integral takes the form

\be \int_{-\infty}^\infty {dx\over s(x)}
\sqrt{w^{(\beta_n,\alpha_n)}(x)}
C_n^{(\beta_n,\alpha_n)}(x) 
\sqrt{w^{(\beta_{n'},\alpha_{n'})}(x)} 
C_{n'}^{(\beta_{n'},\alpha_{n'})}(x)=
\delta_{n\ n'}\ .
\label{orto-1}\ee

As long as the new polynomials $C_n^{(\beta ,\alpha )}(x)$ satisfy
a new hypergeometric equation that generalizes the Jacobi
equation from $s(x)=(1-x^2)$ to $s(x)=(1+x^2)$ we shall refer to the
new polynomials as {\it generalized } classical polynomials.
Equations  (\ref{orth_R}) and  (\ref{orto-1}) show  that the new solutions
have well defined orthogonality properties on the real axes, which
qualifies them as comfortable wave functions in quantum mechanics
applications. 
It is perhaps quite instructive to compare Eq.~(\ref{new_pol})
to the Jacobi equation
\begin{equation}
(1-x^2)\frac{d^2P_n^{(\gamma,\delta)}(x)}{dx^2}
+(\gamma -\delta  -(\gamma +\delta +2)x)\frac{dP_n^{(\gamma,\delta)}(x)}{dx}
-m(m+\gamma +\delta +1)P_n^{(\gamma, \delta )}(x)=0\, .
\label{Jacobi}
\end{equation}
Upon complexification of the argument, $x\to ix$,
the latter equation transforms into
\begin{equation}
(1+x^2)\frac{d^2P_n^{(\gamma,\delta)}(ix)}{dx^2}
+i(\gamma -\delta  -i(\gamma +\delta +2)x)\frac{dP_n^{(\gamma,\delta)}(ix)}{dx}
+m(m+\gamma +\delta +1)P_n^{(\gamma, \delta )}(ix)=0\, .
\label{Jacobi_cplx}
\end{equation}
{}From a formal point of view, Eq.~(\ref{Jacobi_cplx}) can be made
to coincide with  Eq.~(\ref{new_pol}) for the following
parameters:
\begin{equation}
\gamma =\beta -1 -\frac{i\alpha}{2}\, ,
\quad \delta=\beta -1 +\frac{i\alpha}{2}\, .
\label{cplx_csts}
\end{equation}
In this sense one relates in the literature the Jacobi polynomials
of complex arguments and indices to the solutions of the trigonometric
Rosen-Morse potential.
This relation is in our opinion a bit  misleading because the real orthogonal
$C_n^{(\beta ,\alpha)}(x)$ polynomials are apparently a specie that is
fundamentally different from $P_n^{
\left( \beta -1 -\frac{i\alpha}{2},
\beta -1 +\frac{i\alpha }{2}\right)}(ix)$. 

\section{Discussion and concluding remarks}
The physically interesting case of the potential considered here is the one 
of a vanishing $a$ parameter and  the spectrum given in Eq.~(\ref{RVM_spctr}).
It fits perfectly  well the mass splittings of the  
nucleon and  $\Delta $ resonances, results due to 
Refs.~\cite{RVM-KMS},\cite{CK}.
This finding seems to provide a further hint on the possible relevance
of the trigonometric Rosen-Morse potential as an effective QCD confinement
potential that should not be ignored.

Moreover, the exact wave functions of this potential
reveal  quite instructive asymptotic behaviors.
In the zero parameter limit, $a\to 0$, and $b\to 0$, it is easy to show
by explicit calculation that the $R_n (z)$ 
recover the wave functions of the infinite square wall potential,
\be
\lim_{a\to 0,b\to 0}\, R_n(z)=\lim_{a\to 0,b\to 0}
\exp{(-bz)}\sin ^{n+a}C_n^{(\beta_n,\alpha_n)}(\cot z)=
(-1)^{n-1}\sqrt{\frac{2}{\pi}}
\sin nz \, ,
\ee 
which  describe  ``free'' particle motion  within a confining potential.
In that regard one may think of the asymptotically free quarks.

The other instructive asymptotic limit is the one of small $r$
with  $r$ associated with the relative distance between
two particles within the three dimensional version of the potential
in which case Eq.~(\ref{Sch-RMt}) would refer to the radial  part,
$U_n(r)=R_n(r)/r$, of the wave function  and zero angular momentum.
In this case, from Eq.~(\ref{R_z}) one reads off  the ground state
wave function as  
\begin{equation}
 U_1(r)=\frac{R_1(r)}{r}= \exp{\left( -\frac{\alpha_1 r}{2}\right)}
\frac{\sin^{1+a} r}{r}C^{(a,\alpha_1)}_1(\cot r)   \, .
\label{gst_RMt}
\end{equation}
{}For small $r$ and $a=0$ the latter expression 
approaches the ground state of the hydrogen atom,
$ \exp{\left(-\frac{\alpha_1 r}{2}\right)}$,
which would be the nucleon wave function as well if one could
ignore the multi-gluon self interactions and approximate the
three quark problem by a two body quark-diquark one.
This is certainly quite a rough and unrealistic limiting case 
which nonetheless reveals the correct long range
one-gluon exchange mechanism of QCD
as part of the   physical content of the tRM potential.

All in all, the Schr\"odinger equation with
the trigonometric Rosen-Morse potential, besides being an interesting 
quantum mechanics exercise and besides leading to a new
differential equation of mathematical physics that is important
in its own, seems to bear a 
rich information on the quark-gluon dynamics
that may qualify it as an effective QCD potential, a 
possibility that should be kept in mind for future research.

\section*{Acknowledgments}
Work supported by Consejo Nacional de Ciencia y 
Technolog\'ia (CONACyT) Mexico under grant number C01-39280.


\begin{thebibliography}{1}

\bibitem{textbook} 
J.\ J.\ Sakurai, {\it Modern Quantum Mechanics\/}
(Addison-Wesley Pub. Co., Reading 1994);\\
G.\ B.\ Arfken, H.\ J.\ 
Weber, {\it Mathematical Methods for Physicists\/},
 6th ed. (Elsevier-Academic Press, Amsterdam, 2005);\\
I.\ V.\ Kogan,
{\it \/ Problems in Quantum Mechanics}
(Prentice-Hall, Engelwood, 1963);\\
S.\ Fl\"ugge, {\it Practical Quantum Mechanics\/}
(Springer, New York, 1974).
 
\bibitem{handbook}
M. Abramowitz, I. A. Stegun,
\newblock{\em Handbook of Mathematical Functions with Formulas, Graphs and 
Mathematical Tables,} 
\newblock {Dover, 2nd edition, New York, 1972).}



\bibitem{Sukumar} 
C.\  V.\ Sukumar, J.\ Phys.\ A:Math.\ Gen.\ {\bf 18}, 2917 (1998);\\
C.\ V.\ Sukumar, AIP proceedings {\bf 744}, eds. R.\ Bijker et al.,
{\it Supersymmetries in physics and applications},
(New York, 2005), p.\ 167.



\bibitem{Khare} F.\ Cooper, A.\ Khare, U.\ P.\ Sukhatme,
{\it Supersymmetry in Quantum Mechanics}
(World Scientific, Singapore, 2001).


\bibitem{Jacobi-c}
A. B. J. Kuijlaars, A. Martinez-Finkelshtein, R. Orive,
\newblock{\em Orthogonality of Jacobi Polynomials with General Parameters,}
\newblock{E-Print ArXiv: math.CA/0301037 } (2003).

\bibitem{Jacobi-otros}
B. Beckermann, J. Coussement, W. Van Asshe,
\newblock{\em Multiple Wilson and Jacobi-Pi\~neiro Polynomials,}
\newblock{E-Print ArXiv: math.CA/0311055 (2003).}


\bibitem{CK} C.\ B.\ Compean, M.\ Kirchbach,
J.\ Phys.\ A:Math.Gen. {\bf 39}, 547 (2006).











\bibitem{Dennery}
Phylippe Dennery, Andr\'e Krzywicki,
\newblock{\em Mathematics for Physicists} 
\newblock {(Dover,  New York, 1996).}




\bibitem{RVM-KMS}
M. Kirchbach, M. Moshinsky, Yu. F. Smirnov,
\newblock{Phys. Rev. D {\bf 64}, 114005 (2001).}



\end{thebibliography}
\end{document}